\documentstyle[12pt]{article}
\newcommand{\bb}{\begin{equation}}
\newcommand{\ee}{\end{equation}}
\begin{document}
\rightline{hep-th/9603184}
\bigskip
\begin{center}
{\Large{\bf Black Hole Entropy}}\footnote{Invited Talk  delivered
at XVIII IAGRG Conference, Madras, February 1996}\\ \medskip

P. Mitra\footnote{e-mail mitra@tnp.saha.ernet.in}\\
Saha Institute of Nuclear Physics\\
Block AF, Bidhannagar\\
Calcutta 700 064, INDIA
\end{center}
\bigskip
Major  developments  in the history of the subject are critically
reviewed in this talk.
\bigskip\bigskip

\section{Introduction}
A black hole is classically thought of as
a region of intense gravitational field from which
no form of energy -- not even light --
can leak out. The best known example is the Schwarzschild black hole solution
of Einstein's equation. It is described by the metric
\begin{equation}
ds^2=-(1-{2M\over r})dt^2+ (1-{2M\over r})
^{-1}dr^2 +r^2d\Omega^2
\end{equation}
It has a {\it horizon} at $r=2M$, which is a singularity of this coordinate
system, but the curvature is not singular there, and regular {\it Kruskal}
coordinates may be chosen. There is, however, a curvature singularity at
$r=0$, which is to be regarded as the location of a point source of mass $M$.

Another example is the Reissner - Nordstr\"{o}m solution of the Einstein -
Maxwell equations. The metric is given by
\begin{equation}
ds^2=-(1-{2M\over r}+{Q^2\over r^2})dt^2+ (1-{2M\over r}+{Q^2\over r^2})
^{-1}dr^2 +r^2d\Omega^2
\end{equation}
and the electric field by
\bb
F_{tr}={Q\over r^2},
\ee
with $M$ and $Q$ denoting the mass and the charge
respectively. There are apparent singularities at
\begin{equation}
r_\pm=M\pm\sqrt{M^2-Q^2}
\end{equation}
provided $M\ge Q$.
This inequality must hold if a naked singularity is to be avoided
and then there is a horizon at $r_+$.
The limiting case when $Q=M$ and $r_+=r_-$ is referred to as
the extremal case. There is again a curvature singularity at $r=0$, which is
to be regarded as the location of a point source with mass $M$ and charge $Q$.

Classically, a black hole is stable and does not radiate. So a black hole may,
at that level, be assigned zero temperature and correspondingly zero entropy.
But the situation changes when quantum field theory is brought in to describe
the interaction of matter with a black hole background. A careful definition
of the vacuum using regular (Kruskal) coordinates shows that there is a net
flow of particles away from the black hole, with a {\it thermal spectrum}
corresponding to a temperature determined by the black hole parameters like
mass and charge.

The strong gravitational field may be imagined to produce a particle -
antiparticle pair by polarizing the vacuum. If this occurs outside the
horizon, one of the pair may fall in, with the other moving away to infinity,
thus contributing to an outward flow.

If the process continues, it actually becomes faster. This is because a black
hole gets {\it hotter} when it loses mass. The black hole is thus expected to
evaporate completely! A regular spacetime should be left behind, with all the
matter moving away.

This process of evaporation leads to a puzzle. A black hole can be imagined to
start off in a pure state. When it evaporates, there is only thermal
radiation,  which  is  in  a mixed state. Such a transition would
appear to be non-unitary and to involve loss  of  information.  Can it
really occur?

Three different answers have been proposed.\begin{enumerate}
\item Yes. Quantum gravity is non-unitary, and indeed the laws of quantum
mechanics as known to us should be modified to accommodate this kind of
process.
\item No. The thermal nature of the radiation is only an approximation.
There are correlations between different times and the state is really pure.
\item No. The process of evaporation stops at some point to be determined
by an yet unknown theory of quantum gravity. This means that the black hole
leaves a {\it remnant}. \end{enumerate}
It has to be emphasized that each of these proposals involves some new
physics, {\it i.e.,} there is no way of understanding black hole evaporation
in terms of known principles.

It is to be hoped that some day a clear answer will emerge. We shall not
discuss this question any further, but shall pass on to a closely related
topic: black hole entropy. Entropy is, of course, a measure of {\it lack
of information}.

\section{Black hole entropy in the seventies}
A precursor of the idea of entropy in the context of black holes
was the so-called area theorem \cite{Hawk71}. According to this theorem,
the area of the horizon of a system of black holes always increases
in a class of spacetimes. The asymmetry in time is built into the
definition of this class: these spacetimes are predictable from
partial Cauchy hypersurfaces. This result is certainly reminiscent
of thermodynamical entropy.

Some other observations made around that time were collected together
into a set of {\it laws of black hole mechanics} analogous to the
laws of thermodynamics \cite{BCH}.\begin{itemize}
\item The zeroeth law states that the surface gravity $\kappa$
remains constant on the horizon of a black hole.
\item The first law states that
\bb
{\kappa dA\over 8\pi}=dM-\phi dQ,
\ee
where $A$ represents the area of the horizon and $\phi$ the potential
at the horizon.
For the Reissner - Nordstr\"{o}m black hole,
\bb
\kappa={r_+-r_-\over 2r_+^2},~ \phi={Q/r_+},~ A=4\pi r_+^2.
\ee
\item The second law is just the area theorem already stated.
\end{itemize}.

When   these   observations  were  made,  there  was  no  obvious
connection with thermodynamics, it was only a matter of  analogy.
But  it  was  soon  realized  \cite{Bek}  that the existence of a
horizon  imposes  a  limitation  on  the  amount  of  information
available  and hence may lead to an entropy, which should then be
measured by the geometric quantity associated with  the  horizon,
namely its area.
Thus,  upto  a  factor,  $A$  should represent the entropy and ${
\kappa\over 8\pi}$ the temperature.

Not everyone accepted this interpretation of the  laws  of  black
hole  mechanics,  and  in any case the undetermined factor left a
question mark. Fortunately, the problem was solved very soon.  It
was discovered that quantum theory causes dramatic changes in the
behaviour of black hole spacetimes. A scalar field theory in  the
background of a Schwarzschild black hole indicates the occurrence
of radiation of particles \cite{Hawk} at a temperature
\bb
T={\hbar\over 8\pi M}={\hbar\kappa\over 2\pi }.\label{T}
\ee
This demonstrated the  connection  of  the  laws  of  black  hole
mechanics  with  thermodynamics  and  fixed  the scale factor. It
involves Planck's constant and is a quantum effect.

For  a  Schwarzschild  black  hole,  the  first   law   of   {\it
thermodynamics} can be written as
\bb
TdS=dM
\ee
and can be integrated, because of (\ref{T}), to yield
\bb
S={4\pi M^2\over \hbar}={A\over 4 \hbar}.
\ee

Although  the  expression  for $T$ given above is specific to the
case of Schwarzschild  black  holes,  the  relation  between  the
temperature  and  the  surface gravity given in (\ref{T}) is more
generally valid in the case of black holes having $g_{tt}\sim (1-
{r_h\over r})$. The first law of black hole mechanics then becomes
\bb
T d{A\over 4\hbar}=dM-\phi dQ.
\ee
Comparison with the first law of thermodynamics
\bb
{T dS}=dM-\tilde\phi dQ
\ee
is not straightforward because  the  chemical  potential  $\tilde
\phi$  is not clearly known. However, one way of satisfying these
two equations involves the identification
\bb
S={A\over 4\hbar},\qquad \tilde\phi=\phi.\label{std}
\ee

In another approach, the grand partition function is used. For
charged black holes \cite{GH} it
can be related to the classical action by
\begin{equation}
Z_{\rm grand}=e^{-{M- TS-\tilde\phi Q\over T}}\approx e^{-I/\hbar},
\end{equation}
where  the  functional  integral  over  all   configurations   is
semiclassically approximated  by  the  weight factor with  the
classical action $I$. The action is given by a quarter of the area
of the horizon when the Euclidean time goes over one period, {\it
i.e.,} from zero to $1/T$. Consequently,
\bb
M=T(S+{A\over 4\hbar})+\tilde\phi Q.\label{semi}
\ee
Now  there  is a standard formula named after Smarr \cite{Smarr},
\bb
M={\kappa A\over 4\pi}+\phi Q,
\ee
which can be rewritten as
\bb
M=T{A\over 2\hbar}+\phi Q.\label{smarr}
\ee
Comparison with (\ref{semi}) suggests once  again  the  relations
(\ref{std}).  Although the result is the same, it should be noted
that there is a new input: the functional integral.  There  is  a
hope  that  corrections  to the above formulas may be obtained by
improving the  approximation  used  in  the  calculation  of  the
functional integral.

Before closing this section, let us briefly point out that the entropy can be
different from $A/(4\hbar)$ \cite{GMthermo}. The difference between the
first laws of thermodynamics and black hole mechanics can be written as
\bb
TdF=(\phi-\tilde\phi)dQ,
\ee
where $F\equiv S-A/(4\hbar)$. Hence,
\bb
{\partial F\over\partial M}=0,\quad
{\partial F\over\partial Q}= {\phi-\tilde\phi\over T}.\label{F}
\ee
The first equation can be satisfied by an arbitrary function of $Q$,
and the second equation serves to fix $\tilde\phi$ rather than to put
any constraint on $F$.

What happens in the functional integral approach?
The difference between (\ref{semi}) and (\ref{smarr}) is
\bb
TF=(\phi-\tilde\phi)Q.
\ee
By itself this does not impose any restriction but it can be used
to restrict $F$ in conjunction with (\ref{F}). We observe that
\bb
F=Q {\partial F\over\partial Q},
\ee
which means that $F$ is a homogeneous function of $Q$ of degree 1. This
is true even if there are several charges ($Q$ has several components).

\section{Matter around black hole}

The investigation of field theory in the background of a black hole
\cite{Hawk} had given a physical meaning to the temperature of a black hole
but the entropy remained mysterious. An attempt was then made \cite{'tHooft}
to study the entropy of matter in such a background.
It is convenient to employ what is called  the
brick wall boundary condition. Then the
wave function is cut off just outside the horizon at $r_h$. Mathematically,
\begin{equation}
\varphi(x)=0\qquad {\rm at}\;r=r_h+\epsilon
\end{equation}
where   $\epsilon$  is a small, positive, quantity and signifies
an ultraviolet cut-off. There is also an infrared cut-off
\begin{equation}
\varphi(x)=0\qquad {\rm at}\;r=L
\end{equation}
with the box size  $L>>r_h$.

The  wave  equation  for  a scalar  is
\begin{equation}
{1\over\sqrt{-g}}\partial_{\mu}(\sqrt{-g}g^{\mu\nu}\partial_{\nu}\varphi)
-m^2\varphi=0.
\end{equation}
A solution of the form
\begin{equation}
\varphi=e^{-iEt}f_{El}Y_{lm_l}
\end{equation}
satisfies the radial equation
\begin{eqnarray}
-g^{tt}E^2f_{El}&+&{1\over \sqrt{- g}}{\partial\over
\partial r}[\sqrt {-g} g^{rr}{\partial f_{El}\over\partial r}]
-[l(l+1)g^{\theta\theta}+m^2]f_{El}=0.
\end{eqnarray}
An $r$- dependent radial wave number can be introduced from  this
equation by
\begin{eqnarray}
k^2(r,  l,  E)f_{El}&=&-g_{rr}{1\over \sqrt {-g}}{\partial\over
\partial r}[\sqrt {-g} g^{rr}{\partial f_{El}\over\partial r}]
\nonumber\\
&=&g_{rr}[-g^{tt}E^2-l(l+1)g^{\theta\theta} -m^2]f_{El}.
\end{eqnarray}
Only such values of $E$ are to be considered here as make the  above
expression  nonnegative. The values are further restricted by
the semiclassical quantization condition
\begin{equation}
n_r\pi=\int_{r_h+\epsilon}^L~dr~k(r, l, E),
\end{equation}
where $n_r$ has to be a nonnegative integer.

To find the free energy $F$ at inverse temperature $\beta$
one has to sum over states with all possible single- particle
combinations:
\begin{eqnarray}
\beta F&=&\sum_{n_r, l, m_l}\log(1-e^{-\beta E})\nonumber \\
&\approx  &  \int  dl~(2l+1)\int  dn_r\log   (1-e^{-\beta   E})
\nonumber\\
&=&-\int  dl~(2l+1)\int d(\beta E)~(e^{\beta E} -1)^{-1} n_r \nonumber\\
&=& -{\beta\over\pi}\int  dl~(2l+1)
\int dE~(e^{\beta E} -1)^{-1}\int_{r_h+\epsilon}^L
dr~\sqrt{g_{rr}}\nonumber\\
&& \sqrt{-g^{tt}E^2-{l(l+1)}g^{\theta\theta}-m^2}
\nonumber\\
&=& -{2\beta\over 3\pi}\int_{r_h+\epsilon}^L dr~ \sqrt{g_{rr}}
g^{\theta\theta}\nonumber\\&& \int dE~(e^{\beta E} -1)^{-1}
[-g^{tt}E^2-m^2]^{3/2}.
\end{eqnarray}
Here  the  limits  of  integration  for  $l, E$ are such that the
arguments  of  the  square  roots  are   nonnegative.   The   $l$
integration  is  straightforward  and has been explicitly carried
out. The $E$ integral can be evaluated only approximately.

The contribution to the $r$ integral from  large  values  of  $r$
yields  the  expression  for  the free  energy  valid  in  flat
spacetime because of the asymptotic flatness:
\begin{equation}
F_0=-{2\over 9\pi}L^3\int_m^\infty dE{(E^2-m^2)^{3/2} \over
e^{\beta E} -1}.
\end{equation}
This part has to be subtracted out \cite{'tHooft}.
The contribution of the black hole  is  singular  in  the  limit
$\epsilon\to 0$. For ordinary black holes the leading singularity is linear:
\begin{equation}
F\approx -{2\pi^3\over  45\epsilon}
({r_h\over\beta})^4,
\end{equation}
where  the lower limit of the $E$ integral has been approximately
set equal to zero. (There  are
corrections   involving   $m^2\beta^2$   which  will  be  ignored
here.) $\beta$ is now replaced by the reciprocal
of the black hole temperature and the  cutoff $\epsilon$ in $r$
replaced by  one in the ``proper'' radial variable defined by
\begin{equation}
d\tilde r=\sqrt{g_{rr}}dr,
\end{equation}
whence  $\tilde\epsilon \propto   \sqrt{r_h \epsilon }$.
The contribution to the entropy due to the
presence of the black hole is obtained from the formula
\begin{equation}
S=\beta^2 {\partial F\over\partial\beta}
\end{equation}
to be
\begin{equation}
S=  {A\over  360 \pi\tilde\epsilon^2}.
\end{equation}

The appearance of the area led to a lot of interest. The divergence
in the limit of vanishing cutoff $\epsilon$ is clearly due to the
concentration of the matter states near the horizon. The question
then was how the finite result $A/(4[G]\hbar)$ is to be obtained from
this expression. It was suggested that different species of matter,
which have to be summed over, might renormalize the Newton constant
in the denominator of the entropy and might even produce it! However,
an alternative point of view is that the calculation indicated above
refers to the entropy of the matter rather than that of the black hole.
Whereas the calculation of the temperature of matter can tell us about
the temperature of the black hole, the entropy of one need not have any
connection with  that of the other. In support of this  view  one
can  cite the case of {\it extremal} dilatonic black holes, which carry
mass and magnetic charge, but the charge has the maximum  value  that  can
exist   for   a  given  mass  without  giving  rise  to  a  naked
singularity. For these black holes, the entropy as calculated  by
the   above  procedure  is  nonzero,  though  the  area  vanishes
\cite{GM1}.

\section{Counting of states: string based black holes}

While the previous section reviewed an attempt to calculate the entropy
by counting states, it was an entropy in the background of a black hole
rather than the entropy of a black hole itself, which remained unexplained.
Recently, it has been possible to make some progress in this direction,
though  in the context of special black holes arising from string theory
\cite{ashoke}.
In four dimensions the massless bosonic fields of the
heterotic string obtained
by toroidal compactification lead to an effective action with an unbroken
$U(1)^{28}$ gauge symmetry:
\begin{eqnarray}
S&=&{1\over 32\pi}\int d^4x\sqrt{-G}e^{-\Phi}[R_G+G^{\mu\nu}\nabla_{\mu}
\Phi\nabla_{\nu}\Phi+{1\over 8}G^{\mu\nu}Tr(\partial_{\mu}
{\cal M}{\cal L}\partial_{\nu}{\cal M}{\cal L})\nonumber\\
& &-G^{\mu\mu'}G^{\nu\nu'}{F^{( a)}_{\mu\nu}}({\cal L}{\cal M}{\cal L})_{ab}
{F^{( b)}_{{\mu'}{\nu'}}
-{1\over 12}G^{\mu\mu'}G^{\nu\nu'}G^{\rho\rho'}H_{\mu\nu\rho}H_{\mu'\nu'
\rho'}}].
\end{eqnarray}
Here,
${\cal L}=\left(\matrix{-I_{22}& \cr & I_6\cr}\right),$
with $I$ representing an identity matrix,
${\cal  M}$ a symmetric 28 dimensional matrix of scalar fields satisfying
${\cal M\cal L\cal M=\cal L},$
and there are 28 gauge field tensors
$F^{( a)}_{\mu\nu}=\partial_{\mu}A^{( a)}_\nu-
\partial_{\mu}{A^{( a)}_\mu},~~ a=1,...28$
as well as a third rank tensor
$H_{\mu\nu\rho}=\partial_\mu B_{\nu\rho}+2{A^{( a)}_\mu}
{\cal L}_{ab}{F^{( b)}_{\nu\rho}}
+{\rm cyclic~permutations~of~}\mu,\nu,\rho$
corresponding to an antisymmetric tensor field $B$.
The canonical metric defined by
$g_{\mu\nu}=e^{-\Phi}G_{\mu\nu}$
possesses black hole solutions.

The dilaton field is nontrivial, though $H$ still vanishes in the solutions
to be considered. The metric $g_{\mu\nu}$ and the dilaton $\Phi$ are given by
\begin{eqnarray}
ds^2&\equiv& g_{\mu\nu}dx^{\mu}dx^{\nu}\nonumber\\
&=&-{r^2-2mr\over\Delta^{1/2}}dt^2+{\Delta^{1/2}\over
r^2-2mr}dr^2+\Delta^{1/2}d\Omega^2_{II}
\end{eqnarray}
with
\begin{eqnarray}
\Delta &=&r^2\big[r^2+2mr(\cosh\alpha\cosh\gamma-1)+
m^2(\cosh\alpha-\cosh\gamma)^2\big],\nonumber\\{\rm and}
\qquad e^{\Phi}&=&{g^2r^2\over\Delta^{1/2}}.
\end{eqnarray}
Here $\alpha, \gamma$ are real parameters
and $g$ is a constant. The time components of the
gauge fields are given by
\bb
\vec A_t=\cases{-{g\vec n_L\over\sqrt 2}
{mr\sinh\alpha\over\Delta}[r^2\cosh\gamma+mr(\cosh\alpha-
\cosh\gamma)]&$L=1,...   22$\cr -{g\vec n_R\over\sqrt 2}
{mr\sinh\gamma\over\Delta}[r^2\cosh\alpha+mr
(\cosh\alpha-\cosh\gamma)]&$R=23,...   28$}
\ee
with $\vec n_L, \vec n_R$ denoting respectively
22-component and 6-component unit
vectors and
\bb
{\cal M}=I_{28}+\left(\matrix{Pn_Ln_L^T&
Qn_Ln_R^T\cr Qn_Rn_L^T&Pn_Rn_R^T
\cr}\right),
\ee
where
\begin{eqnarray}
P&=&{2m^2r^2\over\Delta}\sinh^2\alpha\sinh^2\gamma\nonumber\\
Q&=&-{2mr\over\Delta}\sinh\alpha\sinh\gamma[r^2+mr(\cosh\alpha\cosh\gamma-1)].
\end{eqnarray}
All other backgrounds vanish for this solution.

The ADM mass of the black hole and its charges are given by
\begin{eqnarray}
M&=&{1\over 4}m(1+\cosh
\alpha\cosh\gamma)\nonumber\\ \vec Q&=&\cases{{g\vec n_L\over\sqrt 2}
m\sinh\alpha\cosh\gamma &$L=1, ...  22$\cr{g\vec
n_R\over\sqrt 2}m\sinh\gamma\cosh\alpha &$R=23, ...  28$}
\end{eqnarray}
The area of the horizon, which is at $r=2m$, is
\bb
{A_H=8\pi m^2(\cosh\alpha+\cosh\gamma),}
\ee
and the inverse temperature (as defined in terms of the surface gravity)
is given by
\bb
{\beta_H=4\pi m(\cosh\alpha+\cosh\gamma).}
\ee
One has to consider the special case
\bb
{m\to 0,\quad\gamma\to\infty,
\quad {\rm with}\quad m\cosh\gamma=m_0,\quad\alpha={\rm finite}.}
\ee
Then
\bb
{\quad A_H=0,\quad T_H={1\over 4\pi m_0},}\label{seventeena}
\ee
and
\bb
{M={m_0\over 4}\cosh\alpha,\quad\vec Q_L=
{gm_0\over\sqrt 2}\sinh\alpha\ \vec n_L,\quad
\vec Q_R={gm_0\over\sqrt 2}\cosh\alpha\ \vec n_R.}\label{nineteen}
\ee
Consequently,
\bb
{M^2={1\over 8g^2}\vec Q_R^2.}
\ee
The black hole is Bogomol'nyi saturated.

This black hole can be identified with a class of massive
string states \cite{duff}, and this is what
allows its entropy to be determined by direct counting.
The density of states in heterotic string theory is given
for a large number $N$ of oscillators by \cite{russo} as
\bb
{\rho\approx{\rm const.}N^{-23/2}e^{2a\sqrt N},}
\ee
where $a_L=2\pi, a_R=\sqrt{2}\pi$.
The numbers of oscillators in the left and right sectors are related to the
mass and charges of
the corresponding states by the usual formula
\bb
{M^2= {g^2\over 8}({\vec Q_L^2\over g^4}+2N_L-2)
= {g^2\over 8}({\vec Q_R^2\over g^4}+2N_R-1).}
\ee
To find the level density in terms of the ADM mass of a black hole, one has
to combine
this formula with the relation between the mass and the charges as applicable
for the solution describing that black hole. Here,
$N_R={1\over 2}$ and the entropy arises from large
values of $N_L$.
\bb
{S=\log\rho\approx 4\pi\sqrt{N_L}\approx{8\pi\over g}
\sqrt{M^2 -{Q_L^2\over 8g^2}}={8\pi\over g\cosh\alpha}M=
{2\pi\over g}m_0.}\label{m_0}
\ee
While this result is nonzero, it must be remembered that in this
limiting case the horizon has a vanishing area. So the area formula
is {\it not} supported. If one desires to express the entropy as
the area of something, one can try to construct  a surface using
various prescriptions \cite{ashoke}.

However, these limiting cases describe special black holes where
the mass and the charges are related with one another, so that these
are {\it extremal cases}. It is known
\cite{HHR} that extremal and nonextremal cases in
the euclidean version are topologically different, so that continuity
need not hold.
If the derivation of an expression for the thermodynamic entropy \cite{GH} is
attempted afresh for the extremal case, with due attention paid to the
fact that {\it the mass and charges are no longer independent as in the
usual cases,} one obtains a form proportional to the mass
of the black hole with an undetermined scale \cite{GM}, as we now show.

Let $\vec\Phi$ represent the chemical potential corresponding to the
charge $\vec Q$.  We can make use of the
$O(22)\times O(6)$ symmetry to write
\bb
\label{dreid}{\vec\Phi=\cases{{\sqrt 2f_L\vec n_L \over 4g}&$L=1,...22$\cr
{\sqrt 2f_R\vec n_R \over 4g}&$R=23,...28$},}
\ee
where $f_L,f_R$ are unknown functions of $m_0$ and $\alpha$.
There are standard expressions for the chemical potential in nonextremal
cases, but we cannot use them for two reasons: first, extremal black holes
may not be continuously connected to nonextremal black holes \cite{HHR}, and
secondly, the standard expressions are calculated by differentiating the mass
with respect to charges at constant {\it area} in the anticipation that
constant area and constant entropy are synonymous, but this is an assumption
we would not like to make.
Only such thermodynamic processes are considered here
which leave the black hole in the class being considered, {\it i.e.,}
all variations  are in the parameters $m_0, \alpha$ and the unit vectors
$\vec n_L,\vec n_R$. Other processes too can occur but are not needed for
this discussion.

Once again, in the leading semiclassical approximation,
the partition function can be taken to be the
exponential of the negative classical action, which vanishes in this case
as the area vanishes.  Hence  the thermodynamic potential
vanishes too and we have, as in (\ref{semi}),
\bb
\label{zweia}{TS=M-\vec\Phi\cdot\vec Q={(\cosh\alpha-f_L\sinh\alpha-
f_R\cosh\alpha)m_0\over 4}.}\ee
Using (\ref{seventeena}), we then have
\bb
\label{zweib}{S=\pi m_0^2(\cosh\alpha-f_L\sinh\alpha-f_R\cosh\alpha).}
\ee
Further, the first law of thermodynamics,
\bb
\label{dreia}{ T dS=dM-\vec\Phi\cdot d\vec Q,}
\ee
takes the form
\bb
\label{zweic}{T{\partial S\over\partial m_0}=
{\partial M\over\partial m_0}-\vec\Phi\cdot{\partial\vec Q\over\partial m_0}
={(\cosh\alpha-f_L\sinh\alpha-f_R\cosh\alpha)\over 4}.}
\ee
This can be written in view of (\ref{zweib}) as
\bb
{{\partial S\over\partial m_0}={S\over m_0},}
\ee
whence
\bb{S=k(\alpha)m_0,}\label{zweie}\ee
with $k(\alpha)$ now an undetermined function of $\alpha$.
This function cannot be fixed by considering the analogue of (\ref{zweic})
where
the $m_0$-derivatives are replaced by $\alpha$-derivatives; what happens
is that $f_L, f_R$ get expressed in terms of $k$.
The string answer (\ref{m_0}) for the entropy is
indeed of the form (\ref{zweie}), with $k(\alpha)$ actually
taking the constant value ${2\pi\over g}$.

\section{Conclusion}
We have come a long way from the seventies, when what seemed like analogues of
the laws of thermodynamics were discovered for black hole physics. First it
became clear that the surface gravity, which was the
analogue of temperature entering those laws, is indeed
proportional to the temperature, the proportionality factor involving Planck's
constant. Then it became apparent that the area of the horizon, which was the
analogue of entropy, naturally enters the expression for the entropy of
matter in the background of a simple black hole. The next step should be to
derive the area expression for the entropy of the black hole itself from a
counting of states. But this would involve a quantum theory of gravity. What
has been achieved in this direction is the embedding of some special black
holes in string theory, leading to a microscopic calculation of the entropy.
This has {\it not} produced the area as the answer, but that is no longer
an occasion for surprise: for
these special black holes the thermodynamical approach also leads to a form
different  from the area but proportional to the mass instead. It is to be
hoped that such calculations will soon be extended to more familiar black
holes.

\bigskip

\noindent {\bf PS}:  For a new development not discussed in this talk, see
\cite{stro}.
\newpage

\end{document}